\documentclass[12pt]{article}
\usepackage{amssymb}
\usepackage{graphicx}
\usepackage{fullpage}
\newtheorem{theorem}{Theorem} 

\newtheorem{conjecture}[theorem]{Conjecture}

\def\eod{\vrule height 10pt width 9pt depth 0pt}

\newcommand{\BBB}[1]{{\mathbb #1}}

\newcommand{\la}{\lambda}

\newcommand{\ve}{\varepsilon}
\newcommand{\Ga}{\Gamma}

\newcommand{\ra}{\rightarrow}
\newcommand{\CC}{{\BBB C}}
\newcommand{\RR}{{\BBB R}}
\newcommand{\DD}{{\BBB D}}
\newcommand{\ZZ}{{\BBB Z}}
\newcommand{\NN}{{\BBB N}}

\newcommand{\beq}{\begin{equation}}
\newcommand{\eeq}{\end{equation}}
\newcommand{\eref}[1]{(\ref{#1})}

\begin{document}
\author{Mark Braverman\thanks{Partially supported by an NSERC 
Post-graduate Scholarship},~~~ Stephen 
Cook\thanks{Partially supported by  an NSERC Discovery grant}\\Department 
of Computer 
Science\\University of Toronto}

\title{Computing over the Reals: Foundations for Scientific Computing}

\maketitle


\medskip


\begin{abstract}
We give a detailed treatment of the ``bit-model'' of computability
and complexity of real functions and subsets of $\RR^n$, and argue
that this is a good way to formalize
many problems of scientific computation.
In Section \ref{intro} we also discuss the alternative Blum-Shub-Smale
model.  In the final section we discuss the issue of whether physical
systems could defeat the Church-Turing Thesis.

\end{abstract}

\section{Introduction}
\label{intro}

The problems of scientific computing often arise from the study of
continuous processes, and questions of computability and complexity
over the reals are of central importance in laying the foundations
for the subject.  The first step is defining a suitable computational
model for functions over the reals.

Computability and complexity over discrete
spaces have been very well studied since the 1930s. 
Different approaches have been
proved to yield equivalent definitions of computability and nearly
equivalent definitions of complexity.  From the tradition of formal
logic we have the notions of recursiveness and Turing machine, and
from computational complexity we have variations on Turing machines
and abstract Random Access Machines (RAMs), which closely model
actual computers.  All of these
converge to define the same well-accepted notion of computability.
The Church-Turing thesis asserts that this formal notion of computability
is broad enough, at least in the discrete setting, to include all
functions that could reasonably construed to be computable.

In the continuous setting, where the objects are numbers in $\RR$,
computability and complexity have received less attention and there
is no one accepted computation model.  Alan Turing
defined the notion of a single computable real number in his landmark
1936 paper \cite{Tur}: a real number is
computable if its decimal expansion can be computed in the discrete
sense (i.e. output by some Turing machine).
But he did not go on to define the notion of
computable real function.  

There are now two main approaches to modeling computation with real number
inputs.  The first approach, which we call the bit-model and which is
the subject of this paper, reflects the fact that computers can only store
finite approximations to real numbers.  Roughly speaking,
a real function $f$ is computable in the bit model if there is an algorithm
which, given a good rational
approximation to $x$, finds a good rational approximation to $f(x)$.

The second approach is the algebraic approach, which abstracts away
the messiness of finite approximations and assumes that real numbers
can be represented exactly and each arithmetic operation can be
performed exactly in one step.  The complexity of a computation is
usually taken to be the number of arithmetic operations (for example
additions and multiplications) performed.  The algebraic approach applies
naturally to arbitrary rings and fields, although for modeling scientific
computation the underlying structure is usually $\RR$ or $\CC$.
Algebraic complexity theory goes back
to the 1950s (see \cite{BorMun,BCS97} for surveys).

For scientific computing the most influential model in the algebraic
setting is due to Blum, Shub and Smale (BSS) \cite{BSS}.
The model and its theory are thoroughly developed in the book \cite{BCSS}
(see also the article \cite{Blum} in the AMS {\em Notices} for an exposition).
In the BSS model, the computer has registers which
can hold arbitrary elements of the underlying ring $R$ and performs
exact arithmetic ($+,-,\cdot$, and $\div$ in the
case $R$ is a field)
and computer programs can branch on conditions based on exact comparisons
between ring elements
($=$, and also $<$, in the case of an ordered field).  Newton's method,
for example, can be nicely presented in the BSS model as a program
(which may not halt)
for finding an approximate zero of a rational function, when $R=\RR$.
A nice feature of the BSS model is its generality: the underlying ring $R$ is
arbitrary, and the resulting computability theory can be studied for each
$R$.  In particular, when $R=\ZZ_2$, the model is equivalent to the
standard bit computer model in the discrete setting.  

One of the big successes of discrete computability theory is
is the {\em un}computability results; especially
the solution of Hilbert's 10th problem \cite{Mat}.  The theorem states that
there is no procedure (e.g. no Turing machine)
which always correctly determines whether
a given Diophantine equation has a solution.
The result is convincing because of general acceptance of the
Church-Turing thesis.

A weakness of the BSS approach as a model of scientific computing
is that uncomputability results do not correspond to
computing practice in the case $R=\RR$.  Since intermediate register values
of a computation are rational functions of the inputs, it is not
hard to see that simple transcendental functions such as $e^x$ are
not explicitly computable by a BSS machine.  In the bit model these functions are computable
because the underlying philosophy is that the inputs and outputs to
the computer are rational approximations to the actual real numbers they
represent.  The definition of computability in the BSS model might
be modified to allow the program to approximate the exact output values,
so that functions like $e^x$ become computable.  However formulating
a good general definition in the BSS model along these lines is not
straightforward: see \cite{Smale97} for an informal treatment and
\cite{Braverman} for a discussion and a possible formal model.

\begin{figure}[!h]
\begin{center}
\includegraphics[angle=0, scale=0.8]{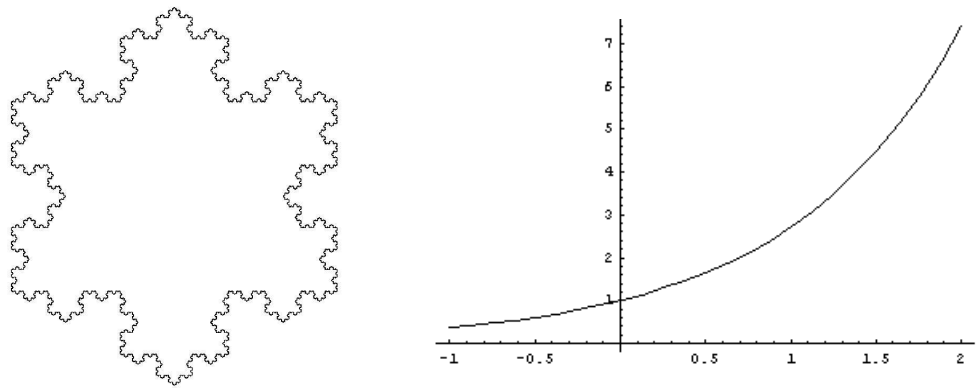}
\end{center}
\caption{
The Koch snowflake and the graph of the equation $y=e^x$
}
\label{koch}
\end{figure}

For uncomputability results, BSS theory concentrates on set decidability
rather than function computation.  A set $C\subseteq \RR^n$ is
{\em decidable} if some BSS computer program halts on each input
$\vec{x}\in\RR^n$ and outputs either 1 or 0, depending on whether
$\vec{x}\in C$.  Theorem 1 in \cite{BCSS} states that if
$C\subseteq \RR^n$ is decided by a BSS program over $\RR$ then
$C$ is a countable disjoint union of semi-algebraic sets.  
A number of sets are proved undecidable as corollaries,
including the Mandelbrot set and all nondegenerate Julia sets
(Fig. \ref{fractimg}).
However again it is hard to interpret
these undecidability results in terms of practical computing,
because simple subsets of $\RR^2$ which can be easily ``drawn'',
such as the Koch snowflake
and the graph of $y=e^x$  (Fig. \ref{koch}) are undecidable
in this sense \cite{Emperor}.

In the bit model there is a nice definition of decidability
(bit-computability) for bounded subsets of $\RR^n$ (Section \ref{s:set-comp}).
For the case of $\RR^2$, the idea is that the set is bit-computable
if some computer program can draw it on a computer screen.  The
program should be able to zoom in at any point in the set and
draw a neighborhood  of that point with arbitrarily fine detail.
Such programs can be easily written
for simple sets such as the Koch snowflake and the graph of the equation
$y=e^x$, and more sophisticated programs can be written for many
Julia sets (Section \ref{s:hard-mand}).  
A Google search on the World Wide Web turns up programs
that apparently do the same for the Mandelbrot set.   However
no one knows how accurate these programs are.  The bit computability
of the Mandelbrot set is an open question, although it holds
subject to a major conjecture (Section \ref{s:hard-mand}).
Because of the Church-Turing thesis, a proof of bit-uncomputability
of the Mandelbrot set would carry some force:  the programs
purporting to draw it must draw it wrong, at least at some level
of detail.

In the rest of the paper we concentrate on the bit model, because
we believe that this is the most accurate abstraction
of how computers are used to solve problems over the reals.
The bit model is not widely appreciated in the mathematical community,
perhaps because its principal references are not written to appeal
to numerical analysts. 
In contrast the BSS model has received widespread attention, partly because
its presentation \cite{BSS}, and especially the
excellent reference \cite{BCSS}, not only present
the model but provide a rigorous
treatment of matters of interest in scientific computation.  The authors
deserve credit not just for presenting an elegant model, but also for
arousing interest in foundational issues for numerical analysis,
and inspiring considerable research.  Undoubtedly the concept of abstracting
away round-off error in computations over the reals poses important
natural questions.  One example is linear programming:  
Although polynomial time algorithms are known for solving linear
programming problems in the bit model,
no such algorithm is known for the BSS model.  The commonly-used
simplex algorithm can be neatly formalized in the BSS model,
but it requires exponential time in the worst case.  It would
be very nice to find a polynomial time BSS algorithm for linear programming.

Here is an outline of the remaining sections.
Section \ref{s:easy-hard} gives examples of easy and hard computational
problems over the reals.  Section \ref{s:bit-model} motivates and
defines the bit model both for computing real functions and
subsets of $\RR^n$.  Computational complexity issues are discussed.
Section \ref{s:hard-mand} treats the computability and complexity
of the Mandelbrot and Julia sets.  Simple programs are available
which seem to compute the Mandelbrot set, but they may draw pieces
which should not be there.  The bit-computability of the Mandelbrot
set is open, but it is implied by a major conjecture.  Many Julia sets
are not only computable, but are efficiently computable (in polynomial time).
On the other hand some Julia sets are uncomputable in a fundamental sense.
Finally Section \ref{s:CTthesis} discusses a fundamental question
related to the Church-Turing thesis:  are there physical systems that
can compute functions which are uncomputable in the standard computer
model?

Some of the material presented here is given in more detail in
\cite{Braverman}.  See \cite{Kobook} and \cite{WeiBook} for
general references on bit-computability models.

\bigskip

\noindent
{\bf \large Acknowledgments}
The authors are grateful to the following people for helpful comments
on a preliminary version of this paper:  Eric Allender, Lenore Blum, Peter Hertling, Ken Jackson,
Charles Rackoff, Klaus Weihrauch, and Michael Yampolsky.

\section{Examples of ``easy" and ``hard" problems}
\label{s:easy-hard}

\subsection{``Easy" problems}

We start with examples of problems over the reals that
should be ``easy" according to any reasonable model. 
The everyday operations, such as addition, subtraction
and multiplication should be considered easy. More generally, 
any of the operations that can be found on a common calculator
can be regarded as ``easy", at least in some reasonable 
region. Such functions include, among 
others, $\sin x$, $e^x$, $\sqrt{x}$, and $\log x$. 

Functions with singularities, such as $x\div y$ and $\tan x$ 
are easily computable on any closed region which excludes 
the singularities. The computational problem usually 
gets harder as we approach the singularity point. 
For example, computing $\tan x$ becomes 
increasingly difficult as $x$ tends to $\frac{\pi}{2}$, because 
the expression becomes increasingly sensitive in $x$.

Some basic numerical problems that are known to have efficient
solutions should also be relatively ``easy" in the model.
This includes inverting a {\em well conditioned} matrix $A$. 
A matrix is well conditioned in this setting if $A^{-1}$ does 
not vary too sharply under small perturbations of $A$. 

Simple problems that arise naturally in the discrete setting should 
usually remain simple in the continuous setting. This includes
problems such as sorting a list of real numbers
and finding shortest paths in a graph with real edge lengths.

When one considers {\em subsets} of $\RR^2$, a set should be considered
``easy", if we can draw it quickly with an arbitrarily high precision. 
Examples include simple ``paintbrush" shapes, such as the ball $B((0,0),2)$
in $\RR^2$, as well as simple fractal sets, such as the Koch 
snowflake (Fig. \ref{koch}).

To summarize, the model should classify a problem as ``easy", if
there is an efficient algorithm to solve it in some practical sense. This algorithm may 
be inspired by a discrete algorithm, a numerical-analytic technique, 
or both.

\subsection{``Hard" problems}

Naturally, the ``hard" problems are the ones for which 
no efficient algorithm exists. For example, it is hard 
to compute an inverse $A^{-1}$ of a poorly conditioned 
matrix $A$. Note that even simple numerical problems, such 
as division $(x+1) \div (y-1)$, become increasingly difficult in the poorly 
conditioned case. It becomes increasingly hard to evaluate the latter expression as 
$y$ approaches $1$. 

Many problems that appear to be computationally hard arise while trying to model 
processes in nature. A famous example is the $N$-body 
problem, which simulates the motion of planets. An even harder
example is solving the Navier-Stokes equations used in simulations for
fluid mechanics. 
We will return to questions of hardness in physical systems in
Section \ref{s:CTthesis}.

Another class of problems that should be hard is natural extensions
of very difficult discrete problems. Consider, for example, the subset sum problem. 
In this problem we are given a list $S = \{a_1, a_2, \ldots, a_n\}$ of numbers and
a number $A$. Our goal is to decide whether there is a subset $\{a_{i_1}, a_{i_2}, \ldots, 
a_{i_k} \}$ that adds up to $A$, i.e. $a_{i_1}+ a_{i_2}+ \ldots + 
a_{i_k} = A$. This problem is widely believed not to have an efficient 
solution in the discrete case. In fact it is $NP$-complete in this 
case (\cite{GareyJohnson}),
and having a polynomial time algorithm for it would imply that
$P=NP$, which is believed to be unlikely.  There is no reason to 
think that it should be any easier in the continuous setting than in the 
discrete case. 

The hardness of numerical problems may significantly vary with the 
domain of application. Consider for example the problem of computing 
the integral $I(x) = \int_{0}^{x} f(t) dt$. While it is very easy to 
compute $I(x)$ from $f(x)$ in the case $f$ is a polynomial, it can 
be extremely difficult in the case $f$ is  a more general efficiently 
computable function. 

\subsection{Quasi-fractal  examples: the Mandelbrot and Julia sets}
\label{MJsets}

The Mandelbrot and the Julia sets are common examples of computer-drawn
sets. Beautiful high-resolution images have become available 
to us with the rapid development of computers. Amazingly, these
extremely complex images arise from very simple discrete iterated 
processes on the complex plane $\CC$. 

For a point $c \in \CC$, define $f_c (z) = z^2+c$. $c$ is 
said to be in the Mandelbrot set $M$ if the iterated sequence 
$c, f_c (c), f_c(f_c(c)), \ldots$ does not diverge to $\infty$. 
While (we believe) very precise images of $M$ can be generated 
on a computer, proving that these images approximate $M$ would 
probably involve solving some difficult open questions about it. 

\begin{figure}[!ht]
\begin{center}
\includegraphics[angle=0, scale=0.65]{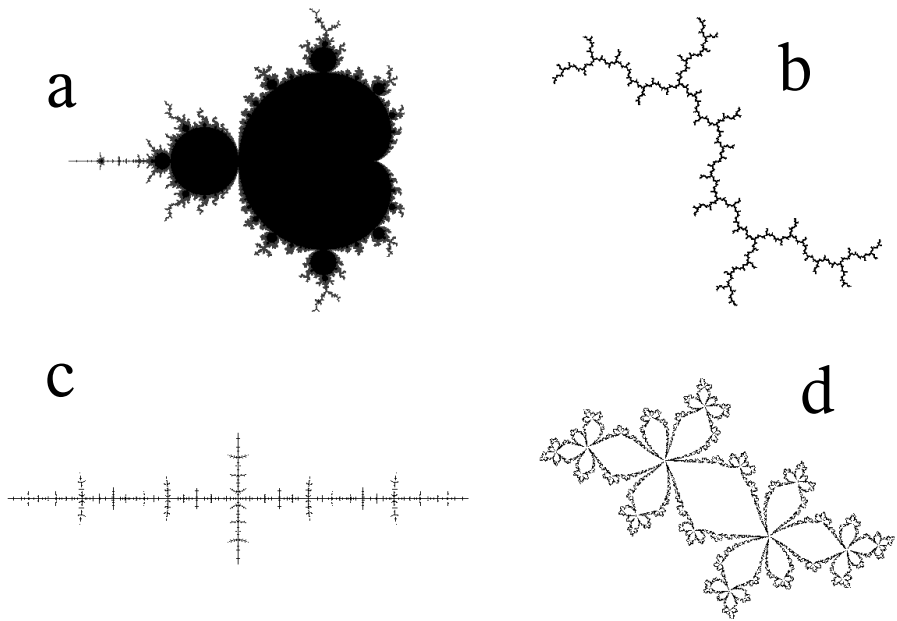}
\end{center}
\caption{{\bf a}: the Mandelbrot set; {\bf b}-{\bf d}: Julia sets with parameter values
of $i$, $-1.57$ and $0.5+0.56i$, respectively.}
\label{fractimg}
\end{figure}

The family of Julia sets is parameterized by functions. In the 
simple case of quadratic functions $f_c (z)=z^2 + c$, the 
{\em filled} Julia set $K_c$ is the set of points $x$, such that 
the sequence $x, f_c(x), f_c (f_c(x)), \ldots$ does not diverge 
to $\infty$. The Julia set $J_c$ is defined as the boundary of 
$K_c$. While many Julia sets, such as the ones in Fig. \ref{fractimg}, are 
quite easy to draw, there are explicit sets of which we
simply cannot produce useful pictures.
 
As we see, it is not {\em a priori} clear whether 
these sets should be ``easy" or ``hard". This gives rise to 
a series of questions: 
\begin{itemize}
\item 
Is the Mandelbrot set computable?
\item 
Which Julia sets are computable?
\item 
Can the Mandelbrot sets and its zoom-ins be drawn quickly on a computer?
\item 
Which Julia sets and their zoom-ins can be drawn quickly on a computer?
\end{itemize}
These questions are meaningless unless we agree upon a model 
of computation for this setting.  In the following sections we
develop such a model, based on ``drawability'' on a computer.

\section{The bit model}
\label{s:bit-model}

\subsection{Bit computability for functions}

The motivation behind the bit model of computation is idealizing 
the process of scientific computing. Consider for example the 
simple task of computing the function $x \mapsto e^x$ for an
$x$ in the interval $[-1,1]$. The most natural solution appears 
to be by taking the Taylor series expansion around $0$:
\beq
\label{s3:e1}
e^x = \sum_{k=0}^{\infty} \frac{x^k}{k!}.
\eeq
Obviously in a practical computation we will only be able 
to add up a finite number of terms from \eref{s3:e1}. 
How many terms should we consider? By taking more terms, 
we can improve the precision of the computation. On the 
other hand, we pay the price of increasing the running time as 
we consider more terms. The answer is that we should take
just enough terms to {\em satisfy our precision needs}.

Depending on the application, we might need the value 
of $e^x$ within different degrees of precision. 
Suppose we are trying to compute it with a precision 
of $2^{-n}$. That is, on an input $x$ we need to output a $y$
such that $|e^x - y| < 2^{-n}$. It suffices to take $n+1$ terms
from \eref{s3:e1} to achieve this level of approximation. 
Indeed, assuming $n\ge 4$, for any $x \in [-1,1]$, 
$$
\left|e^x - \sum_{k=0}^n \frac{x^k}{k!}\right|=
 \left|\sum_{k={n+1}}^{\infty} \frac{x^k}{k!}\right| \le
  \sum_{k={n+1}}^{\infty} \frac{|x^k|}{k!}  \le
 \sum_{k={n+1}}^{\infty} \frac{1}{k!} < 
  \sum_{k={n+1}}^{\infty} \frac{1}{2^{k+1}}  =
 2^{-(n+1)}.
$$
In fact, a smaller number of terms suffices to achieve the
desired precision. We take a portion of the series that 
yields error $2^{-(n+1)}< 2^{-n}$, to allow room for computation 
(round-off) errors in the evaluation of the finite sum. 
All we have to do now is to evaluate the polynomial 
$p_n(x) = \sum_{k=0}^{n} \frac{x^k}{k!}$ within an error of $2^{-(n+1)}$.
To do this, we need to know $x$ with a certain precision
$2^{-m}$.  It is convenient to use {\em dyadic} rationals to express
approximations to $x$ and $e^x$, where the dyadic rationals form the set
$$
    \DD = \{\mbox{$\frac{m}{2^n}$} \mid m\in\ZZ,n\in\NN\}.
$$
We can then take a dyadic $q=\frac{r}{2^s} \in \DD$ such that $|x-q|<2^{-m}$, 
and evaluate $p_n (q)$ within an error of $2^{-(n+2)}$ using finite precision dyadic
arithmetic. This gives us an approximation $y \in \DD$ of $p_n (q)$ such that 
$|y-p_n(q)|<2^{-(n+2)}$. In our example, an 
assumption $|x-q|< 2^{-(n+4)}$ guarantees that $|p_n (x) - p_n(q)|< 2^{-(n+2)}$, 
and we can take $m=n+4$. To summarize, we have 
$$
|e^x - y | \le | e^x - p_n (x)| + | p_n (x) - p_n (q)| +|p_n (q)-y| < 2^{-(n+1)} +
2^{-(n+2)} +2^{-(n+2)} = 2^{-n},
$$
and $y$ is the answer we want. The {\em running time} of the computation 
is dominated by the time it takes to compute our approximation to
$p_n (q)$.
Note that this entire computation is done over the dyadic rationals, and 
can be implemented on a computer.

To find the answer 
we only needed to know $x$ within an error of $2^{-(n+4)}$. This 
is especially important when one tries to compose several computations. 
For example, to compute $e^{e^x-1}$ within en error $2^{-n}$ on 
the appropriate interval, one would need to know $x$ within 
an error of $2^{-(n+8)}$. 

While evaluating the function $e^x$ is hardly a challenge for 
scientific computing, the process described above illustrates 
the main ideas in the bit model of computation. Below are the 
main points that we have seen in this example, and that 
characterize the bit model for computing functions. Suppose 
we are trying to compute $f:[a,b] \ra \RR^n$. Denote the 
program computing $f$ by $M_f$. 
\begin{itemize}
	\item 
	The goal of $M_f (x, n)$ is to compute $f(x)$ within the error of $2^{-n}$;
	\item
	$M_f$ computes a precision parameter $m$, it needs to know $x$ within 
	an error of $2^{-m}$ to proceed with the computation;
	\item 
	$M_f$ receives a dyadic $q = \frac{r}{2^s}$ such that $|x-q|<2^{-m}$;
	\item
	$M_f$ computes a dyadic $y$ such that $|f(x)-y|<2^{-n}$;
	\item 
	the {\em running time} of $M_f (x,n)$ is the time it takes to
	compute $m$ plus the time it takes to compute $y$ from $q$ (both 
	of which have finite representations by bits). 
\end{itemize}

Note that the entire computation of $M_f$ is performed only 
with finitely presented dyadic numbers. There is a nice 
way to present the computation using {\em oracle} terminology. 
An {\em oracle} for a real number $x$ is a function $\phi: \NN \ra \DD$ such 
that for all $n$, $| \phi(n) - x| < 2^{-n}$. Note that the $q$
in the description above can be taken to be $q = \phi(m)$. 
Instead of querying the value of $x$ once, we can allow $M_f$ an 
unlimited access to the oracle $\phi$. The only limitation 
would be that the time it takes $M_f$ to read the output 
of $\phi (m)$ is at least $m$. We emphasize the fact 
that $x$ is presented to $M_f$ as an oracle by writing 
$M_f^\phi (n)$ instead of $M_f (x,n)$. Just as the quality 
of the answer of $M_f (x,n)$ should not depend on the 
specific $2^{-m}$-approximation $q$ for $x$, $M_f^{\phi} (n)$ 
should output a $2^{-n}$-approximation of $f(x)$ for 
{\em any} valid oracle $\phi$ for $x$. 

The running time $T(n)$ of $M_f^{\phi}(n)$ is the worst case
time a computation on this machine can take with 
a valid input and precision $n$. If $T(n)$ is bounded by 
some polynomial $p(n)$, we say that $M_f^\phi$ works
in polynomial time. 

The output of $M_f^{\phi} (n)$ can be viewed itself as 
an oracle for $f(x)$. This allows us to compose functions. 
For example, given a machine $M_f^{\phi}(n)$ for computing $f(x)$,
and a machine $M_g^{\phi}(n)$ for computing $g(x)$, we can 
compute $f \circ g (x)$ by $M_f^{M_g^\phi} (n)$.

This is the bit-computability notion for functions. It was
first proposed by Grzegorczyk \cite{Grz55} and Lacombe \cite{Lac55}. 
It has been since developed and generalized. More recent references on 
the subject include \cite{Kobook}, \cite{PR89}, and \cite{WeiBook}. 
Let us see some examples to illustrate this notion. 

\subsection{Examples of bit-computability}
\label{bit:ex}

We start with a family of the simplest possible functions, the constant 
functions. For a function $f(x)=a$, $a\in \RR$ a constant, we can 
completely ignore the input $x$. The complexity of computing $f(x)$
with precision $2^{-n}$ is the complexity of computing the number $a$ 
within this error.  In the original work by Turing \cite{Tur}, the motivation 
for introducing Turing machines was defining which numbers can 
be computed, and which cannot. 

For example, the numbers $\frac{1}{3}=(0.01010101\ldots)_2$ and $\sqrt{2}$ appear
to be ``easy", with the latter being ``harder" than the former.  There are
also easy algorithms for computing transcendental numbers such as
$\pi$ and $e$.  But there 
are only countably many programs, hence all but countably many reals 
cannot be approximated by any of them. To give a specific example of a
very hard number, consider some standard encoding of all the Diophantine 
equations, $\phi: \{equations\}\ra \NN$. Let $D = \{ \phi(EQ)~:~\mbox{$EQ$ is 
a solvable equation}\} \subset \NN$, and
$$
d = \sum_{n \in D} 4^{-n}.
$$
Then computing $d$ with an arbitrarily high precision would allow 
us to decide whether $\phi(EQ)$ is in $D$ for any specific Diophantine equation 
$EQ$. This would contradict the solution to Hilbert's 10th Problem, which 
states that no such decision procedure can exist (see \cite{Mat}, for example). 


\medskip

The following example will illustrate the bit-computability of a more 
general function. 

\noindent
{\bf Example:} Consider the function $f(x)= \sqrt[3]{1-x^3}$ on the interval $[0,1]$.
It is a composition of two functions: $g:x \mapsto 1-x^3$ and $h:x \mapsto \sqrt[3]{x}$,
both on the $[0,1]$ interval. 
$g$ is easier to compute in this case. To obtain $g(x)$ with precision $2^{-n}$ it 
suffices to query for an approximation $q \in [0,1]$ of $x$ with precision $2^{-(n+2)}$, and 
to return $a=1-q^3$, computed within an error of $2^{-(n+2)}$. Note that the second operation 
is done entirely with bits. We obtain an answer $a$ such that 
$$
|a-g(x)|\le |a-g(q)|+|q^3-x^3|<2^{-(n+2)}+|q-x|\cdot |q^2+q x + x^2| \le
2^{-(n+2)}+2^{-(n+2)}\cdot 3 = 2^{-n}.
$$
The function $h(x)$ is slightly trickier to compute. One possible approach is 
to use Newton's method
to solve (approximately) the equation $\la^3 - x=0$ to
obtain $\la = \sqrt[3]{x}$. 
The fact that $g(x)$ and $h(x)$ are computable is not surprising. In fact, both
these functions can be found on a common scientific calculator. 

In general, all ``calculator functions" are computable on a properly chosen 
domain. For example, $x \mapsto 1/x$ is computable on any domain which excludes 
$0$. We can bound the time required to compute the inverse, if the domain is 
properly bounded away from $0$. 
The only true limiting factor here is that 
computable functions as described above must be continuous on the domain of 
their application. This is because the value of $f(x)$ we compute must be a good approximation 
for {\em all} points near $x$. 

\begin{theorem}
\label{compcont}
Suppose that a function $f: S \ra \RR^k$ is computable. Then $f$ must 
be continuous. 
\end{theorem}

This puts a limitation on the applicability on the computability 
notion above. While it is ``good" at classifying continuous functions, 
it classifies all discontinuous functions, however simple, as being 
uncomputable. We will return to this point later in Section 
\ref{discontsec}.

\subsection{Bit-computability for sets}
\label{s:set-comp}


Just as the bit-computability of functions formalizes finite-precision 
numerical computation, we would like the bit-computability of sets 
to formalize the process of generating images of subsets in $\RR^k$, 
such as the Mandelbrot and Julia sets discussed in Section \ref{MJsets}.

An {\em image} is a collection of {\em pixels}, which can be taken to 
be small balls or cubes of size $\ve$. $\ve$ can be seen as the 
{\em resolution} of the image. The smaller it is, the finer the image 
is. The hardness of producing the image generally increases as $\ve$ 
gets smaller. A collection of pixels $P$ is a good image of a bounded set $S$ if 
the following conditions are fulfilled:
\begin{itemize}
\item 
$P$ covers $S$.  This ensures that we ``see" the
entire set $P$, and
\item
$P$ is contained in the $\ve$-neighborhood of $S$.  This ensures
that we don't get ``irrelevant" components which are far from $S$. 
\end{itemize}
It is convenient to take $\ve = 4\cdot 2^{-n}$ -- our computation 
precision in this case.


Suppose now that we are trying to construct $P$ as a union of 
pixels of radius $2^{-n}$ centered at grid points $(2^{-(n+k)} \cdot \ZZ)^k$.
The basic decision we have to make is whether to draw each particular
pixel or not, so that the union $P$ would satisfy the conditions 
above. To ensure that $P$ covers $S$, we include all the pixels 
that intersect with $S$. To satisfy the second condition, we exclude
all the pixels that are $2^{-n}$-far from $S$. If none of these conditions 
holds, we are in the ``gray" area, and either including or excluding 
a pixel is fine. In other words, we should compute a function 
from the family 
\begin{equation}
\label{setcompeq}
f(d,n) = \left\{ 
\begin{array}{ll}
1 & \mbox{  if  } B(d,2^{-n}) \cap S \ne \emptyset \\
0 & \mbox{  if  } B(d, 2 \cdot 2^{-n}) \cap S = \emptyset \\
0 \mbox{  or  } 1 & \mbox{  otherwise  }
\end{array}
\right.
\end{equation}
for every point $d \in (2^{-(n+k)} \cdot \ZZ)^k$. Here $f$ is computed
in the classical discrete sense.

\noindent
{\bf Example:} A ``simple" set, such as a point, line segment, circle, ellipse, etc. 
is computable if and only if all of its parameters are computable numbers. For example, 
a circle is computable if and only if the coordinates of its center and its radius are
computable.

\smallskip

The way we have arrived at the definition of bit-computability might 
suggest that it is specifically tailored to computer-graphics needs
and is not mathematically robust. This is not the case, as will be seen 
in the following theorem. Recall that the Hausdorff distance between 
bounded subsets of $\RR^k$ is defined as 
$$
d_H(S,T) = \inf\{d~:~S \subset B(T,d), \mbox{ and } T \subset B(S,d)\}. 
$$
We have the following. 
\begin{theorem}
\label{t:3def}
Let $S\subset \RR^k$ be a bounded set. Then the following are equivalent. 
\begin{enumerate}
\item
A function from the family \eref{setcompeq} is computable.
\item
There is a program that for any $\ve>0$ gives an $\ve$-approximation 
of $S$ in the Hausdorff metric.
\item 
The distance function $d_S (x) = \inf\{|x-y| : y \in $S$\}$ is bit-computable. 
\end{enumerate}
{1.} and {3.} remain equivalent even if $S$ is not bounded. 
\end{theorem}

\noindent
{\bf Example:}  The finite 
approximations $K_i$ of the Koch snowflake are  polygons that are obviously
computable.
The convergence $K_i \ra K$ is uniform in the Hausdorff metric. 
So $K$ can be approximated in the Hausdorff metric with any desired 
precision. Thus the Koch snowflake is bit-computable.

\smallskip

The last characterization of set bit-computability in 
Theorem \ref{t:3def} connects the computability 
of sets and functions. 
There is another natural connection
between the computability notions for these objects -- through plots of a function's graph. 

\begin{theorem}
\label{graph:comp}
Let $D \subset \RR^k$ be a closed and bounded computable domain, and 
let $f: D \ra \RR$ be a continuous function. Then $f$ is computable as
a function
if and only if the graph $\Ga_f = \{ (x,f(x)) : x\in D\}$ is computable 
as a set. 
\end{theorem}

\noindent
{\bf Example:} Consider the set $S = \{ (x,y):x,y \in [0,1],~x^3+y^3=1\}$. 
It is the graph of the function $f(x)= \sqrt[3]{1-x^3}$ on $[0,1]$, which 
has been seen to be computable in Section \ref{bit:ex}. By Theorem 
\ref{graph:comp}, $S$ is a bit-computable set. This is despite the 
fact (pointed out in \cite{Blum})
that the only rational points in $S$ are $(0,1)$ and $(1,0)$.

The bit computability notion dates back to  Lacombe \cite{Lac58}. 
We refer the reader to \cite{BW99,WeiBook,Braverman} for a more
detailed discussion.

\subsection{Computational complexity in the bit model}

Since the basic object in the discussion above is a Turing Machine, 
the computational complexity for the bit model follows naturally
from the standard notions of computational complexity in the 
discrete setting. Basically, the time cost of a computation is 
the number of {\em bit} operations required.

For example, the time complexity $T_\pi(n)$ for computing the number 
$\pi$ is the number of bit operations required to compute the 
first $n$ binary digits of a $2^{-n}$-approximation of $\pi$. The time complexity $T_{e^x}(n)$ of 
computing the exponential function $x \mapsto e^x$ on $[-1,1]$, is
the number of bit operations it takes to compute a $2^{-n}$-approximation
of $e^x$ given an $x \in [-1,1]$. This running time is assessed 
at the {\em worst} possible admissible $x$. We have seen that 
both $T_\pi (n)$ and $T_{e^x} (n)$ are bounded by a polynomial in $n$. 

This computational complexity notion can be used to assess the 
hardness of the different numerical-analytic problems arising 
in scientific computing. For example,  the dependence of matrix inversion
hardness on the condition number of the matrix fits nicely into this setting.

Sch\"{o}nhage \cite{Sch1,Sch2} has shown how the fundamental theorem
of algebra can be implemented by a polynomial time algorithm in the
bit model.  More precisely, he has shown how to solve the
following problem in time $O((n^3+n^2s)\log^3(ns))$:
Given any polynomial
$P(z) = a_n z^n + \ldots + a_0$ with $a_j\in\CC$ and
$|P| = \sum_v|a_v| \le 1$, and given any $s\in\NN$, find approximate
linear factors $L_j(z) = u_jz+v_j$ $(1\le j\le n)$ such that
$|P-L_1 L_2\cdots L_n|<2^{-s}$ holds.

The complexity of computing a set is the time $T(n)$ it takes to decide one pixel. 
More formally, it is the time required to compute 
a function from the family \eref{setcompeq}. Thus a set is polynomial-time computable 
if it takes time polynomial in $n$ to decide one pixel of resolution $2^{-n}$. 

To see why this is the ``right" definition, suppose we are trying to 
draw a set $S$ on a computer screen which has a $1000\times 1000$ pixel
resolution. A $2^{-n}$-zoomed in picture of $S$ has $O(2^{2 n})$ pixels
of size $2^{-n}$, and it would take time $O(T(n) \cdot 2^{2 n})$ to 
compute. This quantity is exponential in $n$, even if $T(n)$ is bounded 
by a polynomial. But we are drawing $S$ on a finite-resolution display, 
and we will only need to draw $1000 \cdot 1000 = 10^6$ pixels. Hence 
the running time would be $O(10^6 \cdot T(n)) = O(T(n))$. This running 
 is polynomial in $n$ if and only if $T(n)$ is polynomial. 
Hence $T(n)$ reflects the `true' cost of zooming in, when drawing $S$.

\subsection{Beyond the continuous case 
}
\label{discontsec}

As we have seen earlier, the bit model notion of computability
is very intuitive for sets and for {\em continuous} functions.
However, by Theorem \ref{compcont} it completely excludes 
even the simplest discontinuous functions. Consider for example
the step function 
\beq
\label{step}
s_0 (x) = \left\{ 
\begin{array}{ll}
0, & \mbox{if } x < 0 \\
1, & \mbox{if } x \ge 0 
\end{array}
\right.
\eeq
The function is bit computable on any domain which excludes $0$.
One could make the argument that a physical device really 
{\em cannot} compute $s_0$. There is no bound on the precision 
of $x$ needed to compute $s_0(x)$ near $0$. 
Hence no finite approximation of $x$ suffices to compute $s_0$ even 
within an error of $1/3$. 

On the other hand, one might want to include this function, and other 
simple functions in the computable class, because the primary 
goal of this classification is to distinguish between ``easy"
and ``hard" problems, and computing $s_0$ does not look hard. 
One possibility is to say that a function is computable if 
we can plot its graph. By Theorem \ref{graph:comp}, this definition
extends the standard bit-computability definition in the continuous case. 
It is obvious that is makes $s_0$ computable, since the graph of 
$s_0$ is just a union of two rays. The notion of computational complexity 
can be extended into this more general setting as well. More details
can be found in  \cite{Braverman}.

\section{How hard are the Mandelbrot and Julia sets?}
\label{s:hard-mand}

\begin{figure}[!ht]
\begin{center}
\includegraphics[angle=0, scale=0.8]{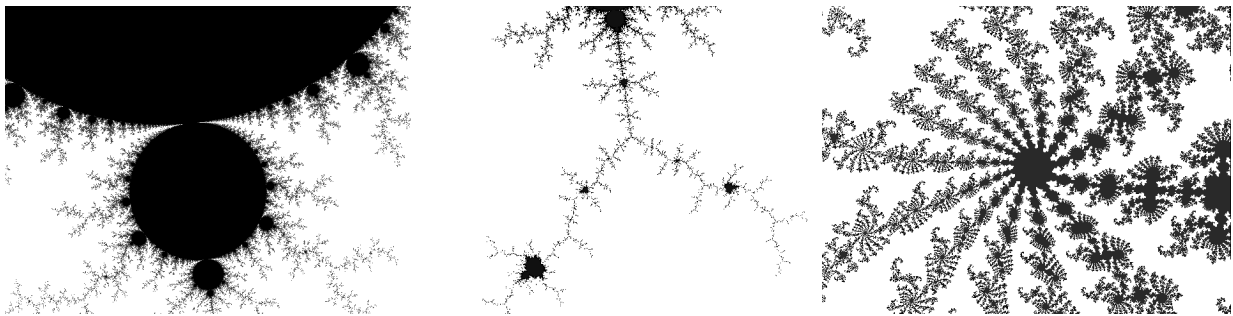}
\end{center}
\caption{A variety of different images arising in zoom-ins of the Mandelbrot set}
\label{mand}
\end{figure}

First let us consider questions of computability of the Mandelbrot set
$M$ (which was defined in Section \ref{MJsets}). Despite the relatively
simple process defining $M$, the set itself is extremely complex, and 
exhibits different kinds of behaviors as we zoom into it. In Fig. 
\ref{mand} we see some of the variety of images arising in $M$. 

\begin{figure}[!ht]
\begin{center}
\includegraphics[angle=0, scale=0.7]{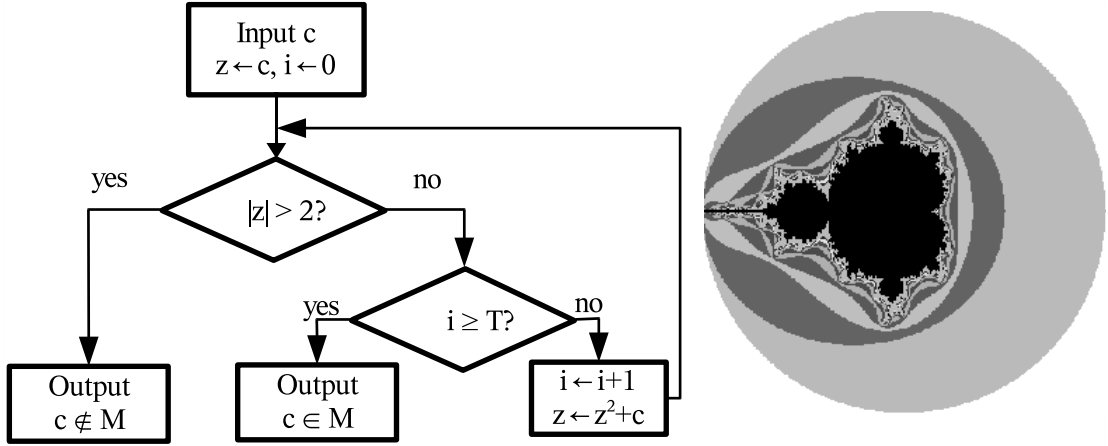}
\end{center}
\caption{The na\"{i}ve algorithm for computing $M$, and some of its outputs}
\label{mandalg}
\end{figure}

The most common algorithm used to compute $M$ is presented on Fig. \ref{mandalg}. The idea
is to fix some number $T$, which is the number of steps for which we are willing 
to iterate. Then for every gridpoint $c$ iterate $f_c (z) = z^2+c$ on $c$ for at most
$T$ steps. If the orbit escapes $B(0,2)$, we know that $c\notin M$.
Otherwise, we say
that $c \in M$. This is equivalent to taking the inverse image of $B(0,2)$ under the 
polynomial map $f^T (c) = \underbrace{f_c \circ f_c \circ \ldots \circ f_c}_{T \mbox{ times}}(c)$.
In Fig. \ref{mandalg} (right) a few of these inverse images and their
convergence to $M$ are shown. 

One problem with this algorithm is that its analysis should take into account
roundoff error involved in the computation $z\leftarrow z^2+c$. 
But there are other problems as well.
For example, we take an arbitrary grid point $c$
to be the representative of an entire pixel. If $c$ is not in $M$, but 
some of its pixel is, we will miss this entire pixel. This problem arises 
especially when we are trying to draw ``thin" components of $M$, such as 
the one in the center  of Fig. \ref{mand}. 
 
Perhaps a deeper objection to this algorithm is the fact that we do not 
have any estimate on the number of steps $T(n)$ we 
need to take to make the picture $2^{-n}$-accurate. That is, a $T(n)$ 
such that for any $c$ which is $2^{-n}$-far from $M$, the orbit of $c$ 
escapes in at most $T(n)$ steps. In fact, no such estimates are known 
in general, and the questions of their existence is equivalent to the 
bit computability of $M$ (cf. \cite{hert}).

Some of the most fundamental properties of $M$ remain open. For 
example, it is conjectured that it is locally connected, but 
with no proof so far.

\begin{conjecture}
\label{locconn}
The Mandelbrot set is locally connected. 
\end{conjecture}

When one looks at a picture of $M$, one sees a somewhat regular 
structure. There is a cardioidal component in the middle, a smaller 
round component to the left of it and two even smaller components
on the sides of the main cardioid. In fact, many of these components 
can be described combinatorially based on the limit behavior of the 
orbit of $c$. E.g., in the main cardioid, the orbit of $c$ converges to 
an attracting point. These components are called {\em hyperbolic components}
because they index the hyperbolic Julia sets that will be discussed 
below. Douady and Hubbard \cite{DH} have shown that Conjecture 
\ref{locconn} 
implies that the interior of $M$ consists entirely of hyperbolic 
components.

\begin{conjecture}
\label{hypcomp}
The interior of $M$ is equal to the union of its hyperbolic components. 
\end{conjecture}

The latter conjecture is known as the {\em Density of Hyperbolicity Conjecture}. 
Hertling \cite{hert} has shown that the Density of Hyperbolicity Conjecture implies 
the computability of $M$. There is a possibility that $M$ is computable 
even without this conjecture holding, but it is hard to imagine 
such a construction without a much deeper understanding of the structure 
of $M$. Moreover, even if $M$ is computable, questions surrounding its 
computational complexity remain wide open.
As we will see, the situation is much clearer for most Julia sets.

A Julia set $J_r$ is defined for every rational function $r$ 
from the Riemann sphere into itself. Here we restrict our attention 
to Julia sets corresponding to quadratic polynomials of the form 
$f_c (z) = z^2+c$. Recall that the filled Julia set $K_c$ is the
set of points $x$ such that the sequence $x, f_c (x), f_c (f_c(x)), \ldots$ 
does not diverge to $\infty$. The Julia set $J_c = \partial K_c$ is the 
boundary of the filled Julia set.

 For every parameter value $c$ there is a different set $J_c$, so 
 in total there are uncountably many Julia sets, and we cannot
 hope to have a machine computing $J_c$ for each value of $c$. 
 What we {\em can} hope for is a machine computing $J_c$ when given
 an access to the parameter $c$ with an arbitrarily high precision. 
 The existence of such a machine and the amount of 
 time the computation takes depends on the properties of 
 the particular Julia set. More information on the properties of 
 Julia sets can be found in \cite{Milnor}.
 
 Computationally, the ``easiest"  case is that of the {\em hyperbolic} Julia sets. 
 These are the sets for which the orbit of the point $0$ either 
 diverges to $\infty$ or converges to an attracting orbit. 
 Equivalently, these are the sets for which there is a smooth 
 metric $\mu$ on a neighborhood $N$ of $J_c$ such that the map $f_c$ is 
 strictly expanding on $N$ in $\mu$. 
  Hence, points escape the neighborhood of $J_c$ exponentially fast. 
  That is, if $d(J_c, x)>2^{-n}$, 
 then the orbit of $x$ will escape some fixed neighborhood of $J_c$ in $O(n)$ steps. This gives us the divergence
 speed estimate we lacked in the computation of the Mandelbrot set $M$,
and shows that in 
 this case $J_c$ is computable in polynomial time (see \cite{RW,Brv04,Ret} 
 for more details). The set $M$ can be viewed as the set of 
 parameters $c$ for which $J_c$ is connected. 
 The hyperbolic Julia sets correspond to the values
 of $c$ which are either outside $M$ or in the hyperbolic components inside $M$. 
 If Conjecture \ref{hypcomp} holds, all the points in the interior of
$M$ correspond to hyperbolic Julia sets as well. None of the points in $\partial M$
correspond to hyperbolic Julia sets. 
 
 We have just seen that the most ``common" Julia sets are computable 
 relatively efficiently. These are the Julia sets that are usually 
 drawn, such as the ones on Fig. \ref{fractimg}. This raises the 
 question of whether {\em all} Julia sets are computable so efficiently. 
 The answer to this question is negative. In fact, it has been shown in 
 \cite{BY} that there are some values of $c$ for which $J_c$ cannot 
 be computed (even with oracle access to $c$).
Moreover, in \cite{BBY} it has been shown that a computable 
 $J_c$ with an explicitly computable $c$ can have an arbitrarily 
 high computational complexity. The 
 constructions are based on Julia sets with Siegel disks. A parameter
  $c$ which ``fools" all the machines attempting to compute $J_c$ is 
 constructed, 
 via a diagonalization similar to the one used in other noncomputability 
results. 
Thus, while ``most" Julia sets are relatively easy to draw, there are some
whose pictures we might never see.

\section{Computational hardness of physical systems and the Church-Turing 
Thesis}
\label{s:CTthesis}

In the previous sections we have developed tools which allow us
to discuss the hardness of computational problem in the continuous 
setting. As in the discrete case, {\em true} hardness of problems
depends on the belief that all physical computational devices have 
roughly the same computational power. 
In this section we present a connection between tractability 
of physical systems in the bit model, and the possibility of 
having computing devices more powerful than the standard computer. 
This provides further motivation for exploring the computability and 
computational complexity of physical problems in the bit model.
The discussion is based in part on \cite{Yao}.

The Church-Turing thesis (CT), in its
common interpretation, is the belief that the Turing machine, which 
is computationally equivalent to the common computer, is the most
general model of computation. That is, if a function can be computed 
using any physical device, then it can be computed by a Turing machine. 

Negative results on in computability theory depend on the 
Church-Turing thesis to give them strong practical meaning.  For 
example, by Hilbert's 10th Problem \cite{Mat},  Diophantine equations cannot be generally 
solved by a Turing Machine. This implies that this problem cannot 
be solved on a standard computer, which is computationally equivalent to 
the Turing Machine. We need the CT to assert that the problem of 
solving these equations cannot be solved on {\em any}
physical device, and thus is {\em truly} hard. 

When we discuss the {\em computational complexity} of problems, 
we are not only interested in whether a function can be computed 
or not, but also in the time it would take to compute a function. 
The Extended Church-Turing thesis (ECT) states that any physical system 
is roughly as efficient as a Turing machine. That is, if it computed a 
function $f$ in time $T(n)$, then $f$ can be computed by a Turing Machine 
in time $T(n)^c$ for some constant $c$. 

In recent years, the ECT has been questioned, in particular by 
advancements in the theory of {\em quantum computation}. In principle,
if a quantum computer could be implemented, it would allow us 
to factor an integer $N$ in time polynomial in $\log N$ \cite{shor}. This 
would probably violate the ECT, since factoring is believed to require 
superpolynomial time  on a classical computer.  On the other hand
there is no apparent way in which quantum computation would violate the CT. 

\begin{figure}[!ht]
\begin{center}
\includegraphics[angle=0, scale=0.8]{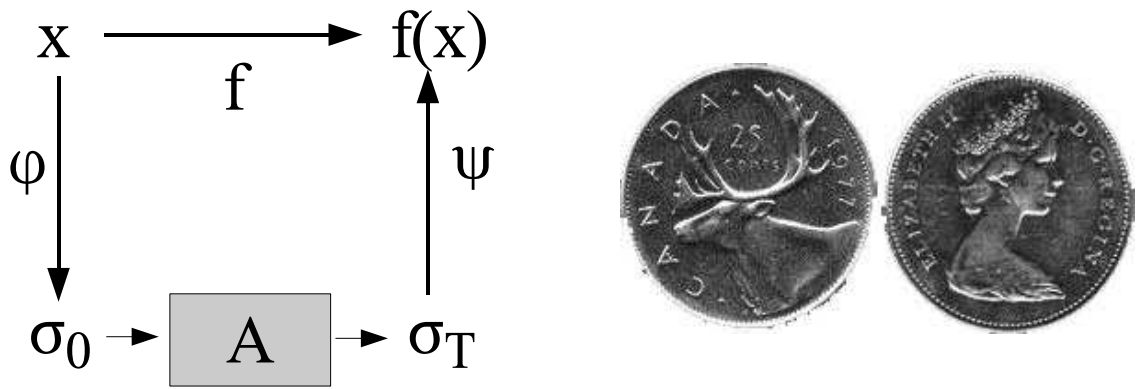}
\end{center}
\caption{Computing $f$ using a ``hard" physical device $A$ (left); a fair coin cannot 
be considered a ``hard" device}
\label{physical}
\end{figure}

Let $f$ be some uncomputable function. 
Suppose that we had a physical system $A$ and two computable 
translators $\phi$ and $\psi$, such that $\phi$ translates an 
input $x$ to $f$ into a state $\phi(x)$ of $A$. The evolution
of $A$ on $\sigma_0 = \phi(x)$ should yield a state $\sigma_T$ such that 
$\psi(\sigma_T) = f(x)$ (Fig. \ref{physical}). This means that at least in principle 
we should be able to build a physical device which would allow 
us to compute an uncomputable function. To compute $f$ on an input $x$, we 
translate $x$ into a state $\sigma_0 = \phi(x)$ of $A$.  
We then allow $A$ to evolve from state $\sigma_0$ to $\sigma_T$ -- this 
is the part of the computation that cannot be simulated by a computer. 
We translate $\sigma_T$ to obtain the output $f(x)=\psi(\sigma_T)$. 

To make this scheme practical, we should require
  $A$ to be {\em robust} around $\sigma_0=\phi(x)$, at least in
some probabilistic sense. That is, for a small random perturbation
 $\sigma_0 + \ve$ of $\sigma_0$, $\psi(\sigma_T)$ should be equal 
 to $f(x)$ with high probability.
 
 It is apparent from this discussion that such an $A$ should be 
 hard to simulate numerically, for otherwise $f$ would be computable
 via a numerical simulation of $A$. On the other hand, ``hardness
 to simulate" does not immediately imply ``computational hardness". 
 Consider for example a fair coin. It is virtually impossible to 
 simulate a coin toss numerically due to the extreme sensitivity
 of the process to small changes in initial conditions. Despite
 its unpredictability, a fair coin toss cannot be used to compute ``hard"
 functions, because it is not robust. In fact, due to noise, for 
 any initial conditions which puts the coin far enough from the 
 ground, we {\em know} the probability distribution of the outcome: 50\% ``heads"
 and 50\% ``tails". Another example where ``theoretical hardness" of 
 the wave equation does not immediately imply a violation of the CT is 
 presented in \cite{Waves}.
 
 This leads to a question that is essentially equivalent to the CT:
 
 \smallskip
 \centerline{  ($*$) { \em Is there a {\em robust} physical system that is 
 hard to simulate numerically?}}
\smallskip

This is a question that can be formulated  in the framework
of bit-computability. Since the only numerical simulations 
a computer can perform are bit simulations, hardness of some robust 
system $A$ for the bit model implies a positive answer for $(*)$. 
On the other hand, proving that all computationally hard systems 
are inherently highly unstable would yield a negative answer to this 
question.

Note that even if $(*)$ has a positive answer and CT does not hold,
and there exists some physical 
device $A$ that can compute an uncomputable function $f$, it does not imply that
this device could serve some ``useful" purpose. That is, it might be 
able to compute some meaningless function $f$, but not any of the 
``interesting" undecidable problems, such as solving Diophantine 
equations or the halting problem.


\begin{thebibliography}{abc}











\bibitem[BBY05]{BBY} I. Binder, M. Braverman, M. Yampolsky. 
{On computational complexity of Siegel Julia sets,} 
{\em arXiv.org e-Print archive}, 2005.
Available from \verb+http://www.arxiv.org/abs/math.DS/0502354+
 

\bibitem[Blum04]{Blum}
L. Blum, Computing over the Reals:  Where Turing Meets Newton. 
{\em Notices of the Amer. Math. Soc} {\bf 51}, 1024-1034, 2004.


\bibitem[BSS89]{BSS}
L. Blum, M. Shub, S. Smale, On a theory of computation and complexity
over the real numbers: NP-completeness, recursive functions and universal
machines. {\em Bulletin of the Amer. Math. Soc.} {\bf 21}, 1-46, 1989.

\bibitem[BCSS98]{BCSS}
L. Blum, F. Cucker, M. Shub, S. Smale, Complexity and Real Computation, 
Springer, New York, 1998.


\bibitem[BM75]{BorMun}
A. Borodin, I. Munro, The Computational Complexity of Algebraic and
Numeric Problems, Elsevier, New York, 1975.





\bibitem[Brt03]{Emperor}
V. Brattka, The  emperor's new recursiveness: The epigraph of the 
exponential  function in two models of computability. In Masami Ito and 
Teruo Imaoka,  editors, {\em Words, Languages \& Combinatorics III}, pp. 
63-72, Singapore, 2003. 
World Scientific Publishing. {\em ICWLC 2000}, Kyoto, Japan, March 14-18, 2000.

\bibitem[BW99]{BW99}
V. Brattka, K. Weihrauch, Computability of Subsets of Euclidean 
Space I: Closed and Compact Subsets, {\em Theoretical Computer Science},
{\bf 219}, pp. 65-93, 1999. 



\bibitem[Brv04]{Brv04}
M. Braverman, Hyperbolic Julia Sets are Poly-Time Computable. Proc. of 
CCA 2004, in ENTCS, vol {\bf 120}, pp. 17-30.

\bibitem[BY04]{BY} M. Braverman, M. Yampolsky. {Non-computable Julia sets,} 
{\em arXiv.org e-Print archive}, 2004.
Available from \verb+http://www.arxiv.org/abs/math.DS/0406416+

\bibitem[Brv05]{Braverman}
M. Braverman, On the Complexity of Real Functions.
{\em arXiv.org e-Print archive}, 2005.
Available from \verb+http://www.arxiv.org/abs/cs.CC/0502066+.
Accepted for 46th Annual IEEE Symposium on Foundations of Computer Science
(FOCS 2005).

\bibitem[BCS97]{BCS97}
P. Burgisser, M. Clausen, M. A. Shokrollahi, Algebraic Complexity Theory.
Springer, New York, 1997.





\bibitem[DH82]{DH}
A.Douadyand, J.H.Hubbard,
It\'eration des polynomes quadratiques complexes. 
{\it C. R. Acad. Sci. Paris,} {\bf 294},  pp. 123-126, 1982.

\bibitem[GJ79]{GareyJohnson}
M.R. Garey, D.S. Johnson,
Computers and Intractability:  A Guide to the Theory of NP-Completeness,
W.H.Freeman and Company, 1979.

\bibitem[Grz55]{Grz55}
A. Grzegorczyk, Computable functionals, {\em Fund. Math.} {\bf 42}, pp. 168-202, 
1955. 

\bibitem[Hert05]{hert} P. Hertling, Is the Mandelbrot set computable? {\em Math. Logic Quart.} 
{\bf 51}, issue 1, pp. 5-18, 2005.



\bibitem[Ko91]{Kobook}
K. Ko, Complexity Theory of Real Functions, Birkh\"{a}user, Boston, 1991.


\bibitem[Lac55]{Lac55}
D. Lacombe, Classes r\'{e}cursivement   ferm\'es et fonctions 
majorantes, {\em C. R. Acad. Sci. Paris}, {\bf 240}, pp. 716-718, 1955. 

\bibitem[Lac58]{Lac58}
D. Lacombe, Les ensembles r\'{e}cursivement ouverts ou ferm\'es, et leurs 
applications \`a l'Analyse r\'ecursive, {\em C. R. Acad. Sci. 
Paris}, {\bf 246}, pp. 28-31, 1958. 

\bibitem[Mat93]{Mat} Y. Matiyasevich, {\it Hilbert's Tenth Problem}, The MIT Press, Cambridge, London, 
1993. 



\bibitem[Mil00]{Milnor}
J. Milnor, Dynamics in One Complex Variable - Introductory
Lectures, second edition, Vieweg, 2000.



\bibitem[PR89]{PR89}
M.B. Pour-El, J.I. Richards, Computability in Analysis and Physics, 
Perspectives in Mathematical Logic, Springer, Berlin, 1989.



\bibitem[Ret04]{Ret}
R. Rettinger, A Fast Algorithm for Julia Sets of Hyperbolic Rational Functions. Proc. of 
CCA 2004, in ENTCS, vol {\bf 120}, pp. 145-157. 

\bibitem[RW03]{RW}
R. Rettinger, K. Weihrauch, The Computational Complexity of Some Julia 
Sets, in {\em STOC'03}, June 9-11, 2003, San Diego, California, USA. 

\bibitem[Sch82]{Sch1}
A. Sch\"{o}nhage, The fundamental theorem of algebra in terms of
computational complexity, Technical report, Math. Institut der. Univ. 
T\"{u}bingen, 1982.

\bibitem[Sch87]{Sch2}
A. Sch\"{o}hage, Equation solving in terms of computational complexity.
In {\em Proceedings of the International Congress of Mathematicians, 1986},
A. Gleason, Ed., Amer. Math. Soc., 1987.

\bibitem[Shor97]{shor} P. Shor, Polynomial-time algorithms for prime factorization and discrete 
logarithms on a quantum computer. SIAM J. Comput. {\bf 26}, pp. 1484-1509, 1997.

\bibitem[Sma97]{Smale97} S. Smale, Complexity theory and numerical analysis.
{\em Acta Numerica}, {\bf 6}, pp. 523-551, 1997.







\bibitem[Tur36]{Tur} A. M. Turing, On Computable Numbers, With an 
Application to the 
Entscheidungsproblem. In {\it Proceedings, London Mathematical Society}, 
1936, pp. 230-265.

\bibitem[Wei00]{WeiBook}
K. Weihrauch, Computable Analysis, Springer, Berlin, 2000.

\bibitem[WZ02]{Waves}
K. Weihrauch, N. Zhong, Is wave propagation computable
  or can wave computers beat the Turing machine?
  {\em Proc. London Math Soc.} (3) {\bf 85} pp. 312-332, 2002.


\bibitem[Yao02]{Yao}
A. Yao, Classical Physics and the Church-Turing Thesis, Electronic Colloquium
on Computational Complexity, Report No. {\bf 62}, 2002. 


\end{thebibliography}
 \end{document}